# Bandwidth-Efficient Synchronization for Fiber Optic Transmission: System Performance Measurements

*Oluyemi Omomukuyo, Octavia A. Dobre, Ramachandran Venkatesan, and Telex M. N. Ngatched*

By any measure, the various services and applications which are crucial to today's society rely on fast, efficient, and reliable information exchange. Nowadays, most of this information traffic is carried over long distances by optical fiber, which has intrinsic advantages such as wide transmission bandwidth and low attenuation. However, continuing traffic growth has imposed many challenges, especially for equipment manufacturers who have to develop optical transmission solutions to handle the demand for higher data rates without incurring increased capital and operational costs. A feasible approach to overcoming these challenges is to scale the channel capacity by employing orthogonal frequency division multiplexing (OFDM) super-channels. However, OFDM is sensitive to synchronization errors, which can result in complete failure of the receiver-based digital signal processing. Measurement results of various existing OFDM synchronization methods have highlighted inherent limitations with regards to poor system performance, which determines the quality-of-service level perceived by the end user, and complexity, which throws doubts as to their suitability for implementation in actual equipment. In this article, we first provide a brief overview of optical transmission systems and some of their performance specifications. We then present a simple, robust, and bandwidth-efficient OFDM synchronization method, and carry out measurements to validate the presented synchronization method with the aid of an experimental setup.

**Introduction**

In its most basic form, an optical transmission system, as shown in Fig. 1a, consists of an optical source (usually a laser), optical fiber, an optical detector (usually a photodiode), as well as various other optical components such as modulators between the optical source and detector. In this system, some sort of information, encoded in the form of an electrical signal, is used to modulate the intensity of an optical carrier either directly using the optical source, or indirectly via an external modulator. The modulated optical signal travels along the optical fiber, where it is directly detected by the optical detector and converted back to an electrical signal. Such a simple system is referred to as an intensity-modulated direct detection (IMDD) system.

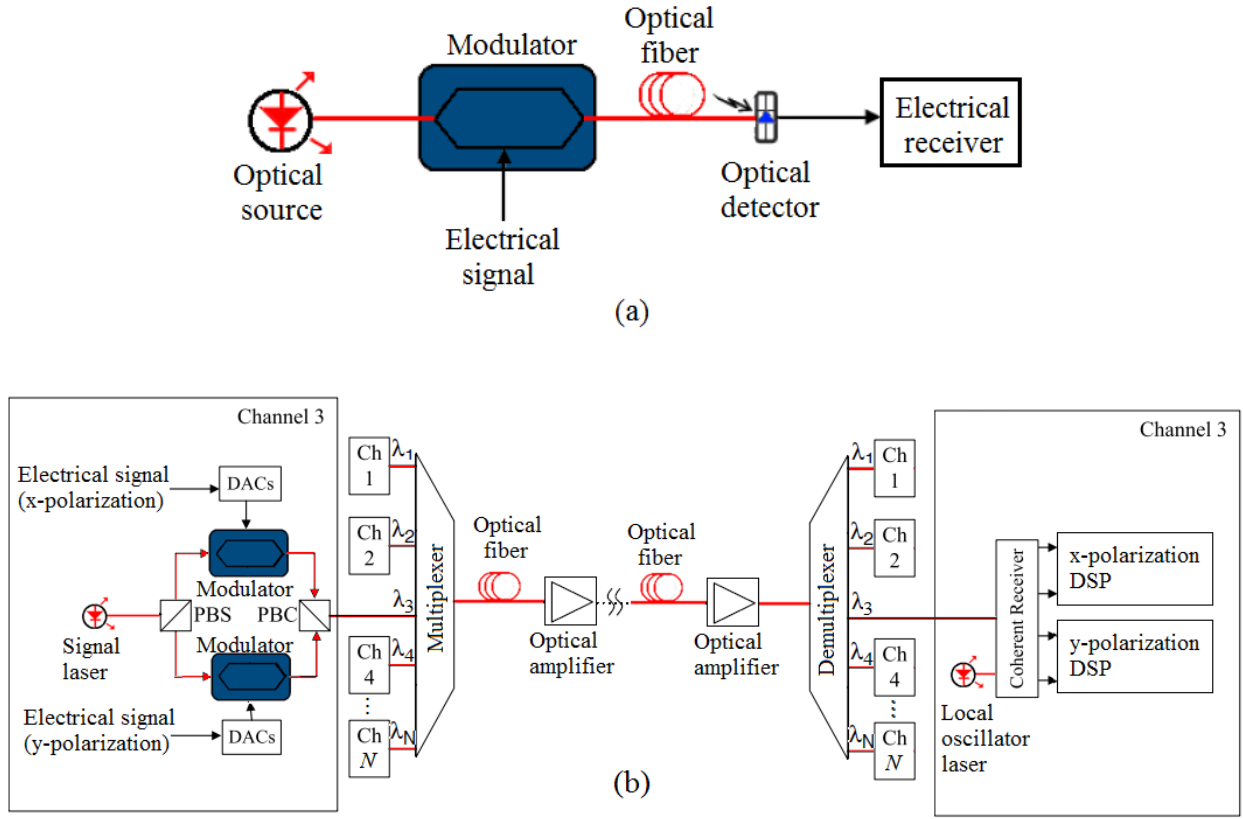

*Fig. 1.* (a) Basic IMDD transmission system employing an external modulator; (b) Optical transmission system featuring PDM, WDM, and coherent detection. DAC: digital-to-analog converter. PBS: polarization beam splitter. PBC: polarization beam combiner. Ch: channel.

IMDD systems are limited by low transmission data rate (measured in bits per second), and are typically used for short distances. In order to increase the transmission data rate and reach, modern optical transmission systems usually employ the following techniques:

- Coherent detection, where the modulated optical signal is combined coherently with another optical signal at the receiver. The deployment of commercial coherent detection optical systems has been facilitated by rapid advances in digital signal processing (DSP).
- Polarization division multiplexing (PDM), which is based on the ability of light, as an electromagnetic wave, to oscillate with more than one orientation. By launching optical signals into the fiber on two orthogonal states of polarization (*x* and *y*) at an identical wavelength, PDM results in doubling the data rate.

- Wavelength division multiplexing (WDM), where several data channels are modulated onto optical carriers at different wavelengths, multiplexed, and then launched into the optical fiber. These channels are demultiplexed at the receiver.

Fig. 1b depicts the detailed architecture of an optical transmission system employing the three techniques mentioned above. The increase in transmission data rate and reach from using coherent optical systems instead of IMDD systems comes at the cost of increased complexity. In addition, performance impairments at high data rates become significant, such as the fiber chromatic dispersion, polarization mode dispersion (PMD), and amplified spontaneous emission (ASE) noise generated by the optical amplifiers. Consequently, characterizing these impairments, and carrying out measurements using specialized instruments is very essential for performance assessment, testing, and troubleshooting.

**Measurement in Optical Transmission Systems**

As in any communication system, the bit error rate (BER) is the fundamental parameter that quantifies the fidelity of data transmission. The BER is a ratio of the number of bits received in error to the total number of received bits. For high-speed long-haul optical transmission systems, the major sources of bit errors are the ASE noise, and signal distortion induced by linear and non-linear fiber impairments. In such optical systems, a BER of 1e-15, achieved with forward error correction, is often required.

The main linear fiber impairment is dispersion, which occurs when a light pulse is spread out during fiber transmission, interfering with neighboring pulses, and resulting in errors. Non-linear impairments increase in significance exponentially with an increase in the power launched into the fiber, putting a limit on the achievable transmission reach. Modern coherent optical systems utilize DSP at the receiver to compensate for these impairments. System performance measurements are thus crucial in ensuring that these DSP algorithms can tolerate the level of optical impairments that the system will be exposed to. For equipment manufacturers, these measurement results can be exploited to quantify the quality-of-service level perceived by end users [1]. In this article, we highlight two of the most important optical measurements usually performed.

*Optical Signal-to-Noise Ratio Measurements*

For long-haul optical transmission systems that utilize several in-line optical amplifiers as shown in Fig. 1b, the optical signal-to-noise ratio (OSNR) is an important parameter that indicates the degree of impairment brought about by ASE noise. The OSNR is the ratio of the signal optical power to the ASE noise power, and optical receivers are usually specified in terms of the required OSNR (R-OSNR), which is the OSNR level required to obtain a target BER. The OSNR penalty, obtained as the ratio between the actual OSNR (at which the target BER is achieved) and the receiver R-OSNR, is a useful measurement for evaluating the robustness of a DSP algorithm to various impairments. In experimental setups, noise loading is a technique which is used to vary the OSNR in order to evaluate system performance in the presence of ASE noise. This article describes the noise loading technique in more detail later on.

*PMD Measurements*

PMD arises because of the non-circular symmetry of actual fiber brought about by external stress as well as imperfections introduced during manufacturing. This asymmetry results in a difference in propagation speeds through the fiber of the two orthogonal polarization states. PMD becomes increasingly significant at high data rates, and as such, PMD measurements using specialized laboratory instruments are used to evaluate the PMD tolerance. One such instrument is the PMD emulator that generates varying levels of PMD-induced delay so as to accurately and rapidly reproduce the variations of PMD in actual fiber. We have made use of a PMD emulator for the experimental validation of the presented synchronization method.

**Scaling Channel Capacity in Optical Transmission**

Modern society demands increased bandwidth to support a plethora of applications and services such as video streaming, social media, and cloud computing, all of which have contributed to an exponential growth in data traffic. This growth is also being accelerated by mobile access. Despite the increase in transmission data rate brought about by utilizing WDM and PDM in current commercial coherent optical systems, a pertinent question presents itself, which is:

*How can this traffic growth be sustained without substantially increasing the cost per bit?*

Providing an answer to this question is very crucial for optical equipment manufacturers, and current research and development efforts are centered on scaling the channel capacity of next-generation coherent optical transmission systems without increasing the operational complexity or the cost per bit per Hz [2]. A feasible solution is to employ optical super-channels, where the spacing between conventional WDM channels is reduced, improving the bandwidth efficiency (in bits per second per Hz). These super-channels are then routed together over a common optical link, as shown in Fig. 2.

Coherent optical OFDM (CO-OFDM) [3] is one of the enabling technologies used in super-channel coherent optical systems. OFDM makes use of several low-rate sinc-shaped subcarriers which are orthogonally multiplexed, as shown in Fig. 2. This orthogonality is what permits the WDM channel spacing to be reduced without suffering crosstalk penalties. However, OFDM is more sensitive to synchronization errors when compared with single carrier modulation schemes [4]. These synchronization errors can result in loss of orthogonality and significant crosstalk in a CO-OFDM super-channel. Synchronization is therefore crucial for practical CO-OFDM receivers. The three key synchronization subsystems are:

- Frame synchronization, which is concerned with estimating the correct start of each OFDM symbol in the received frame.
- Frequency synchronization, whose functionality is compensating for the carrier frequency offset (CFO) caused by the incoherence of the lasers at the transmitter and receiver.
- Sampling clock synchronization, which is concerned with estimating and compensating for the mismatch between the sampling clocks of the digital-to-analog converter (DAC) in the transmitter and the analog-to-digital converter (ADC) in the receiver.

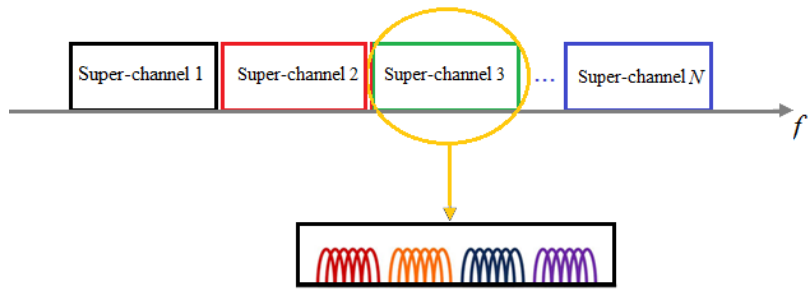

*Fig. 2.* CO-OFDM super-channels made up of tightly-packed WDM channels. Each WDM channel comprises several sinc-shaped subcarriers.

Many synchronization methods in the literature are limited by one or more of the following: poor frame synchronization performance, limited CFO estimation range, and complexity. Furthermore, because these methods require different training symbols for the different synchronization tasks, see e.g. [5], the net bit rate is reduced, which reduces the bandwidth efficiency.

In this article, we present a simple, robust, and bandwidth-efficient method to overcome all the above-mentioned limitations without sacrificing performance, power consumption or cost. Furthermore, with the aid of experimental measurements, we then discuss and evaluate the performance of the method in a PDM CO-OFDM system.

**Operation Principle of Synchronization Method**

To achieve a bandwidth-efficient synchronization solution, we use the same training symbols to carry out joint frame, frequency, and sampling clock synchronization. Golay complementary sequences (GCS) [6], which can achieve minimum mean square error performance of channel estimation, have been used as the training symbols. These training symbols are arranged in the Alamouti scheme [7], where there are two training symbols per polarization.

*Frame Synchronization*

The training symbols $A$ and $B$, of length $N_{sc}$, where $N_{sc}$ is the number of OFDM subcarriers, are zero-padded with $N - N_{sc}$ zeros, where $N$ is the inverse fast Fourier transform (IFFT) size. The starting point of the frame for the $x$-polarization, $\hat{d}_x$, can be determined by utilizing the autocorrelative property of the GCS to construct a timing metric [8]:

$$\hat{d}_x = arg\left\{max_d\left[\frac{\left|\sum_{n=0}^{N+N_{cp}-1}\left(P_x^A(d) - P_x^B(d)\right)\right|^2}{\left(2\sum_{n=0}^{N+N_{cp}-1}|r_x(d+n)|^2\right)^2}\right]\right\}, \qquad (1)$$

with

$$P_x^A(d) = r_x(d+n)p(n) \times r_y\left(d + \alpha + N + 2N_{cp} + mod\left([N_{cp} - n], N\right)\right),$$
$$P_x^B(d) = r_y(d + \alpha + n)p(n) \times r_x\left(d + N + 2N_{cp} + mod\left([N_{cp} - n], N\right)\right),$$

where $r_x(d)$ and $r_y(d)$ are the discrete received samples for the $x$ and $y$ polarizations, respectively, $\alpha$ is the PMD-induced delay (in samples), $N_{cp}$ is the number of cyclic prefix samples, $mod(.)$ is the modulo operator, and $p(n) \in (-1,1]$, with $n = 0,1,\cdots, N + N_{cp} - 1$, is a pseudo-random sequence used to ensure the timing metric is impulse-shaped with a distinct peak. The interested reader is referred to [8] for similar equations for the y-polarization.

*Frequency Synchronization*

The frequency synchronization is divided into two sub-stages: fractional frequency synchronization, for estimating the fractional part of the CFO with a magnitude $\leq \Delta f$, and integer frequency synchronization, for estimating the integer part of the CFO, where $\Delta f$ is the OFDM subcarrier spacing. After obtaining the right start of the CO-OFDM frame, the estimate of the fractional CFO is obtained from (1) by setting $\alpha = 0$ and $d = \hat{d}_x$, as explained in detail in [8]. Once the fractional CFO has been compensated for, the integer CFO is estimated by performing a cross-correlation between the frequency domain samples of the first received training symbol and the sum of the samples of the original training symbols. The CFO, obtained by summing the integer and fractional CFO estimates, is then compensated for by multiplying the received time domain signal samples by an inverse rotation factor [9].

*Sampling Clock Synchronization*

The sampling clock synchronization is based on the principle that the sampling clock offset (SCO) induces a phase rotation which varies linearly with both the OFDM subcarrier and symbol indices [10], [11]. The estimate of the relative SCO, $\hat{\gamma}$, between the clocks of the DAC and ADC is [10]:

$$\hat{\gamma} = \frac{mN}{2\pi s_B (N + N_{cp})}, \qquad (2)$$

where $s_B$ is the OFDM symbol index for $B$ (which is 2 in this case), and $m$ represents the SCO-induced phase shift slope, which is obtained from the phase angles of the complex division of the received samples of $B$ by their corresponding transmitted samples. The SCO estimate is fed-back to a time-domain interpolation, where the SCO is compensated for by resampling. Any residual SCO can be eliminated by repeating the feedback loop.

**Measurement Setup and Results**

In order to verify the synchronization method, PDM 16-QAM measurements have been conducted on a CO-OFDM system using an experimental testbed, whose schematic is shown in Fig. 3.

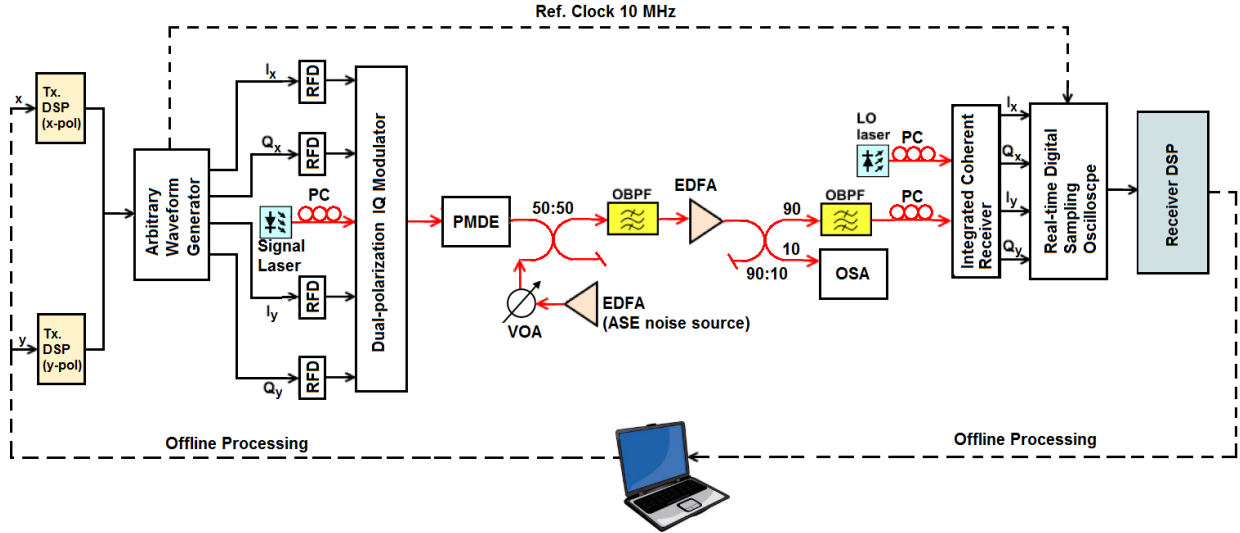

*Fig. 3.* Experimental setup. RFD: RF driver, PMDE: polarization mode dispersion emulator, PC: polarization controller, EDFA: erbium-doped fiber amplifier, OSA: optical spectrum analyzer, OBPF: optical band-pass filter, VOA: variable optical attenuator, LO: local oscillator.

For the experimental demonstrations, the measurable parameters we have focused on include timing errors, CFO, and SCO-induced phase offset, which have been evaluated using performance metrics such as BER, OSNR penalty, mean squared error (MSE), and constellation measurements. In the experimental system, the DSP procedures for both transmitter and receiver are carried out offline in MATLAB. At the transmitter, a pseudo-random bit sequence is mapped onto the OFDM subcarriers into parallel complex data using 16-QAM modulation. The time-domain signal is then generated using a 512-point IFFT with a 46-sample cyclic prefix. Out of the 512 IFFT channels, 416 are data subcarriers, 85 are unmodulated subcarriers used for oversampling, one unmodulated subcarrier is reserved for the DC term, and the remaining 10 are unmodulated subcarriers centered around DC. The training symbols are inserted at the beginning of the OFDM frame to carry out frame synchronization.

The generated CO-OFDM signal is saved as a single-column ASCII text file and then loaded into the volatile memory of a Tektronix AWG70002A Arbitrary Waveform Generator (AWG) operating at a sampling rate of 25 GSa/s and 9-bit DAC vertical resolution. The analog in-phase and quadrature-phase signals for both polarizations are then amplified by a 32-Gbaud RF driver with 20-dB gain. These amplified signals are then used as the electrical drives to a null-biased 30-GHz bandwidth dual-polarization IQ modulator. The optical source to the IQ modulator is an integrated tunable laser assembly with linewidth of 10 kHz and center emission wavelength of 1550 nm. A PMD emulator is employed to investigate the performance of the presented method in the presence of the PMD-induced delay.

For the OSNR measurements, the noise loading is carried out by using an erbium-doped fiber amplifier (EDFA) as an ASE noise source. This EDFA is operated without an input, and its output is coupled with the output of the PMD emulator using a 50:50 coupler. The level of the ASE noise added, and consequently, the OSNR level, is varied by adjusting the attenuation of a variable optical attenuator connected to the ASE noise source. The generated optical signal is then filtered by a 0.6-nm optical band-pass filter to suppress the out-of-band ASE noise, before being amplified by a second EDFA with 15-dB gain. The output of the second EDFA is split into two paths using a 90:10 coupler. The 10% path of the coupler is fed into an optical spectrum analyzer to monitor the OSNR. The 90% path is filtered by another optical band-pass filter, before being detected by a 32-Gbaud integrated coherent receiver with a 10-kHz linewidth local oscillator. The wavelength of the local oscillator is tuned as appropriate to yield the desired amount of CFO for the frequency synchronization measurements.

The coherently-detected signal is then captured by a Tektronix DPO 72304DX, 23-GHz, real-time digital sampling oscilloscope (DSO) operated at a sampling rate of 50 GS/s with 8-bit ADC resolution. In order to introduce SCO for sampling clock synchronization measurements, a 10-MHz output reference clock from the AWG is connected to the DSO. This enables the sampling clocks of the DAC and ADC to be synchronized, allowing for flexibility in emulating various amounts of SCO by simply adjusting the sampling rate of the AWG. The horizontal position of the DSO is varied to emulate different time delays for the frame synchronization measurements.

*Discussion of Measurement Results*

Fig. 4a shows the measured timing and frequency metrics of the method for a CFO of 2.5 GHz and an SCO of 160 ppm at an OSNR of 15 dB, respectively. It can be seen that both metrics are impulse-shaped, and yield a distinct peak at the right values of the timing and frequency offsets, even in the presence of significant CFO and SCO.

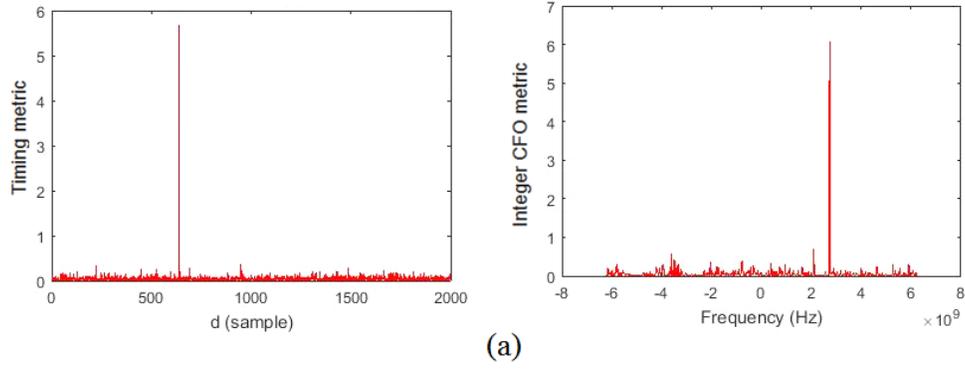

(a)

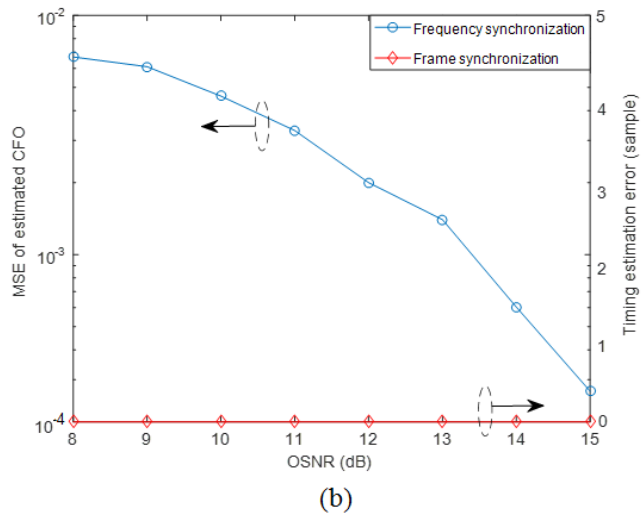

(b)

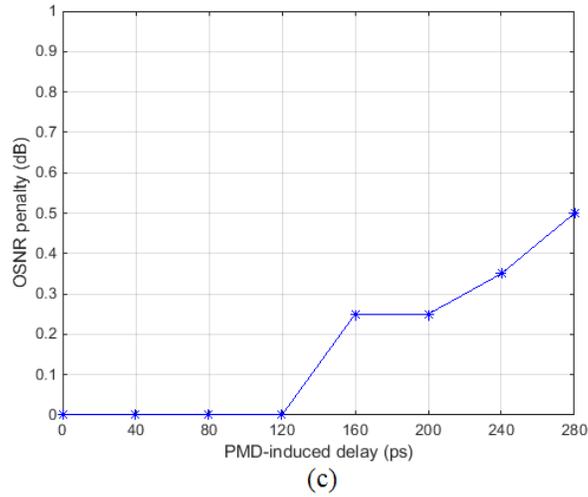

(c)

*Fig. 4.* (a) Measured metrics for the x-polarization at 15-dB OSNR (left: timing metric, right: integer CFO metric); (b) Frame and frequency synchronization performance in the presence of ASE noise; (c) OSNR penalty in the presence of PMD-induced delay.

The results of Fig. 4b demonstrate the robustness of the frame and frequency synchronization to ASE noise, with no frame synchronization errors observed, and a maximum CFO MSE of 6.7e-3 (corresponding to a CFO estimation error of ~4 MHz) observed in all cases. Fig. 4(c) shows the OSNR penalty, based on the R-OSNR to achieve a target BER of 1.8e-2 (supported with FEC), as a function of the PMD-induced delay. The results show the robustness of the method to PMD, with a maximum OSNR penalty of 0.5 dB obtained for a PMD-induced delay of up to 280 ps.

Fig. 5a shows the actual and least-squares fit of the phase rotation caused by an SCO of 160 ppm, and for a CFO of 2.5 GHz, confirming that the SCO-induced phase rotation varies linearly with the OFDM subcarrier index. From the least-squares fit, the slope can be obtained, and the SCO estimated using (2). Fig. 5b shows the measured 16-QAM constellations after channel equalization without, and with SCO compensation for an SCO of 160 ppm.

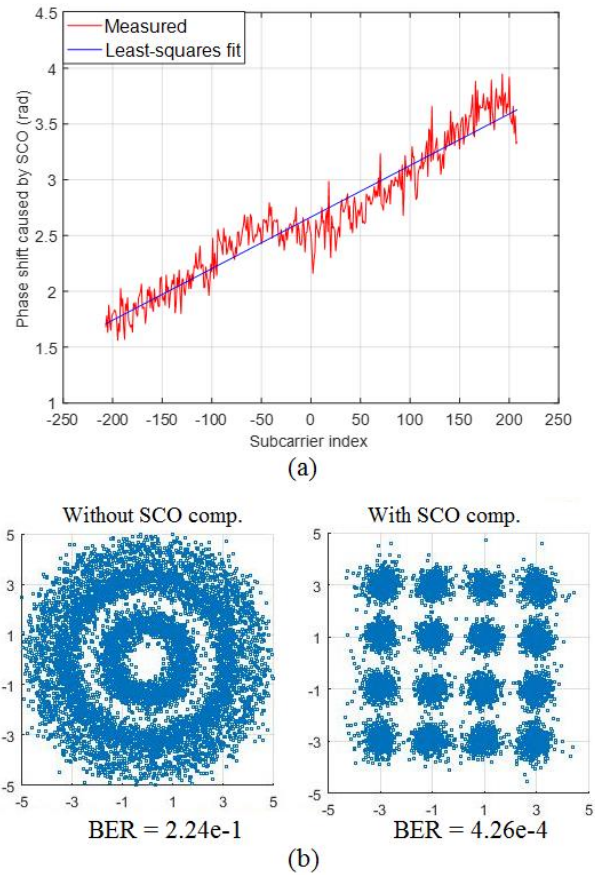

*Fig. 5.* (a) SCO-induced phase rotation versus subcarrier index; (b) measured constellations for an SCO of 160 ppm: without and with SCO compensation.

It can be seen from the constellation that when no mechanism for SCO compensation is in place, the system suffers significant performance degradation from the SCO-induced phase rotation, resulting in a poor BER of 2.24e-1. When the SCO is compensated using the method, the system performance improves remarkably, with the BER dropping to 4.26e-4.

**Conclusion**

Optical fiber communications has revolutionized the entire telecommunications industry, and technological advances to date have allowed the commercial deployment of high-speed long-haul optical transmission systems. Given the continued growth of data traffic, these technologies are fast approaching their practical limits. CO-OFDM super-channels offer a feasible solution to this problem without significantly increasing cost. In this article, we have provided an overview of optical transmission systems and some of the important optical measurement parameters and techniques, which play a vital role in performance assessment, troubleshooting and improvement. Through measurements, we then experimentally demonstrated a bandwidth-efficient synchronization method for a CO-OFDM system. The measurements performed on the system demonstrate the robustness of the method to ASE noise and PMD, where no timing errors have been observed, and accurate CFO and SCO estimation are achieved.

**References**

[1] L. Angrisani, D. Petri, and M. Yeary, "Instrumentation and measurement in communication systems," *IEEE Instrum. Meas. Mag.*, vol. 18, no.2, pp. 4–10, Apr. 2015.
[2] Optical Internetworking Forum, "Technology options for 400G implementation," *OIF-Tech-Options-400G-01.0* (July 2015).
[3] W. Shieh, H. Bao, and Y. Tang, "Coherent optical OFDM: theory and design," *Opt. Express*, vol. 16, no. 2, pp. 841–859, Jan. 2008.
[4] F. He and L. Wu, "Timing and ranging models based on OFDM synchronization," in *Proc. IEEE Int. Instrum. Meas. Technol. Conf. (I2MTC)*, pp. 1-6, 2007.
[5] O. Omomukuyo *et al.*, "Joint timing and frequency synchronization based on weighted CAZAC sequences for reduced-guard-interval CO-OFDM systems," *Opt. Express*, vol. 23, no. 5, pp. 5777-5788, Mar. 2015.
[6] M. Golay, "Complementary series," *IRE Trans. Inf. Theory*, vol. IT-7, no. 2, pp. 82-82, Apr. 1961.
[7] S. Alamouti, "A simple transmit diversity technique for wireless communications," *IEEE J. Sel. Areas Commun*. vol. 16, no. 8, pp. 1451-1458, Oct. 1998.

[8] O. Omomukuyo *et al.*, "Robust frame and frequency synchronization based on Alamouti coding for RGI-CO-OFDM," *IEEE Photon. Techn. Lett.*, vol. 28, no. 24, pp. 2783-2786, Dec. 2016.

[9] J.-W. Kim *et al.*, "Adaptive carrier frequency offset compensation for OFDMA systems," in *Proc. IEEE Int. Instrum. Meas. Technol. Conf. (I2MTC)*, pp. 1-4, 2012.

[10] O. Omomukuyo *et al.*, "Simple sampling clock synchronization scheme for reduced-guard-interval coherent optical OFDM systems," *IET Elect. Lett*, vol. 51, no. 24, pp. 2026-2028, Nov. 2015.

[11] A. Makki *et al.*, "Indoor localization using 802.11 time differences of arrival," *IEEE Trans. Instrum. Meas.*, vol. 65, no. 3, pp. 614-623, Mar. 2016.